\documentclass[aps,prl,floatfix, preprint]{revtex4}
\usepackage{graphicx}

\begin{document}

\title{Anisotropic Spin Diffusion in Trapped Boltzmann Gases}

\author{W. J. Mullin$^{1}$}
\author{R. J. Ragan$^{2}$} 
\affiliation{$^{1}$Physics Department, Hasbrouck
Laboratory, University of Massachusetts, Amherst, MA 01003}
\affiliation{$^{2}$Physics Department, University of Wiscosin at Lacrosse,
La Crosse, WI 54601} 

\date{\today }

\begin{abstract}
\noindent Recent experiments in a mixture of two hyperfine states of trapped
Bose gases show behavior analogous to a spin-1/2 system, including
transverse spin waves and other familiar Leggett-Rice-type effects. We have
derived the kinetic equations applicable to these systems, including the
spin dependence of interparticle interactions in the collision integral, and
have solved for spin-wave frequencies and longitudinal and transverse
diffusion constants in the Boltzmann limit. We find that, while the
transverse and longitudinal collision times for trapped Fermi gases are
identical, the Bose gas shows diffusion anisotropy. Moreover, the lack of
spin isotropy in the interactions leads to the non-conservation of
transverse spin, which in turn has novel effects on the hydrodynamic modes.
\end{abstract}

\maketitle



In recent JILA\ experiments,\cite{JILA},\cite{JILA2} a mixture of two
hyperfine states was found to segregate by species. The theoretical
explanation\cite{Levitov}-\cite{Brad} for this behavior is based on the two states
playing the role of a pseudo-spin-1/2 system, having transverse spin waves.
The theory of these new effects is based on old ideas of the transport
properties of polarized homogeneous quantum gases of real spins, such as $%
^{3}$He gas and solutions of $^{3}$He in liquid $^{4}$He,\cite{cand},\cite
{MullJeon} transcribed to the trapped gas pseudo-spin case.

Besides spin waves, the theory for homogeneous polarized fermions or bosons
led to the prediction of anisotropic spin diffusion in the degenerate state.%
\cite{cand},\cite{MullJeon},\cite{JeonGlyde} When a spin nonuniformity
is longitudinal, that is, with a variation in the magnitude of the
magnetization, the spin diffusion coefficient is $D_{\parallel }$.  On
the other hand, in a spin-echo experiment, the magnitude of the
magnetization is uniform but it varies spatially in direction.  The
corresponding diffusion coefficient, $D_{\perp \;}$ is less than
$D_{\parallel }$ when the system is polarized and degenerate. 
Experimentally this feature has been seen, but was not always in
reasonable accord with theory.\cite{cand} Moreover, Fomin\cite{Fomin}
has suggested the effect should not exist.  However, a recent
experiment\cite{cand} has overcome several possible experimental
objections and finds good agreement with theory.  Moreover Mineev has
very recently presented theoretical analysis that questions the
validity of Fomin's approach.\cite{Mineev}

Thus it seems useful to see whether a similar difference between
longitudinal and transverse diffusion in trapped gases might provide
an alternative testing ground for this question.  However, what we
show here is that the physical possibility of having differing
interaction parameters between up-up, down-down, and up-down states
(interaction anisotropy) provides a new physical basis for anisotropic
spin diffusion for bosons even in the Boltzmann limit.\cite{footnote}
For longitudinal diffusion in the Boltzmann limit only up-down
scattering contributes.  However, in the transverse case, two spins at
differing angles approach one another, and the scattering can be
analyzed as being a superposition of, say, up-up and up-down
scattering.  In the fermion s-wave case, the up-up part gives no
contribution, and, in the Boltzmann limit, the diffusion coefficients
are identical.  In that case one must go to the degenerate limit to
see the anisotropy, which then is expected to arise because the
density of scattering states differs in longitudinal and transverse
cases.\cite{MullJeon} On the other hand, for bosons, for which both
the up-up and down-down scattering rates do contribute, we find an
anisotropy even in the Boltzmann limit, but only if the various
scattering lengths differ. We have here the striking effect that, 
although both gases obey Boltzmann statistics, there is a macroscopic 
difference between fermion and boson behavior.

The presence of interaction anisotropy provides another unusual
effect, namely that transverse spin is not conserved.\cite{Nikuni2}
This leads to a decay of the transverse spin (a $T_{2}$ process) that
seriously affects the hydrodynamic modes of the system.  Below we
first use the moments method to compute the spectra of the
lowest-lying longitudinal and transverse modes.  However, with that
method we obtain a transverse decay rate $\gamma_{\perp}$ that
diverges as $\tau$ approaches zero, in contrast to the usual diffusive
behavior where $\gamma_{\perp} \propto \tau$.  In this case it is
necessary to solve the \emph{local} hydrodynamic equations to find the
correct behavior, in which the hydrodynamic solutions are
localized at the low-density regions at the edges of the cloud where
the collision time is longer. The result is a much smaller decay rate
than that obtained with the moments method.

In our previous work, Ref.~\onlinecite{MullJeon}, we derived an analog of
the Landau-Silin equation for a $2\times 2$ density operator $\widehat{n}%
_{p}$ (here acting in the pseudo-spin space), with effective
mean-field single particle energy matrix $\hat{\epsilon}_{p}$.  We can
write the density and single-particle energy in a Pauli
representation as $\widehat{n}_{p}=\frac{1}{2}\left( f_{p}\hat{I}+\mathbf{m}%
_{p}\cdot \hat{\mathbf{\sigma}}\right) $ and $\widehat{\varepsilon }%
_{p}=\left( e_{p}\hat{I}+\mathbf{h}_{p}\cdot \mathbf{\hat{\sigma}}\right) $%
where $\mathbf{\hat{\sigma}}$ is a Pauli matrix, $\frac{1}{2}(f_{p}\pm
m_{pz})$ give the diagonal components of the density
$n_{pi}=n_{pii}$, while $\mathbf{m}_{p}$ represents the polarization,
which in equilibrium is along the axis $\mathbf{\hat{z}}$.  We find
the following approximate equation for $\mathbf{m}_{p}$:
\begin{equation}
\frac{\partial \mathbf{m}_{p}}{\partial t}-\frac{2}{\hbar }\mathbf{h\times m}%
_{p}+\sum_{i}\left[ \frac{p_{i}}{m}\frac{\partial \mathbf{m}_{p}}{\partial %
r_{i}}-\frac{\partial U}{\partial r_{i}}\frac{\partial \mathbf{m}_{p}}{%
\partial p_{i}}\right] =\mathrm{Tr}\left\{ \mathbf{\hat{\sigma}}\hat{I}%
_{p}\right\} ~  \label{mmeq}
\end{equation}
with $m_{pz}(\mathbf{r})=n_{p1}-n_{p2}$ and 
$n_{p12}(\mathbf{r})=n_{p21}^{*}=\frac{1}{2}%
m_{p-}(\mathbf{r})=\frac{1}{2}(m_{px}-im_{py})$.
The $2\times 2$ collision integral is $\hat{I}_{p}.$ The effective mean
magnetic field 
\begin{equation}
\mathbf{h}=\frac{\hbar \Omega _{0}}{2}%
\mathbf{\hat{z}}+\eta \frac{t_{12}}{2}\mathbf{M}
\end{equation}
where 
$\hbar \Omega_{0}=V_{1}-V_{2}+\left[ 
\left(t_{11}-t_{12}\right)n_{1}
-\left( t_{22}-t_{12}\right) n_{2}\right]\nonumber\\\times\left(1+\eta \right)
$. In these $\eta $ is $1$ ($-1$) for bosons (fermions); $\mathbf{M}(\mathbf{r}%
)=\int d\mathbf{p}/h^{3}\;\mathbf{m}_{p}(\mathbf{r});$ $n_{i}(\mathbf{r}%
)=\int d\mathbf{p}/h^{3}\;n_{p}(\mathbf{r});$ $V_{i}$ is the external
field for species $i$; $U=\frac{1}{2}(V_{1}+V_{2})$; and
$M_{z}=n_{1}-n_{2}$.  The $t$'s can be evaluated in terms of the
measured scattering lengths $%
a_{\alpha \beta }$ by using $t_{\alpha \beta }=4\pi \hbar a_{\alpha \beta
}/m.$ 

The equilibrium solution in the Boltzmann limit is 
$m_{p}^{(0)}=\mathcal{M}(\beta \hbar \bar{%
\omega})^{3}\exp[-\beta (p^{2}/2m+U)]$
where $N$ is the total number of particles, $N_{i}$ is the number of species 
$i,$ $\mathcal{M}=N_{1}-N_{2}$ is the total magnetization, and $\bar{\omega}%
\equiv (\omega _{x}\omega _{y}\omega _{z})^{1/3}.$

We have derived the collision integral for the Boltzmann case when the
various interaction paramenters differ. Our expression agrees with the same
quantity derived in Refs.~\onlinecite{Nikuni2} and \onlinecite {Brad}, and
reduces properly to previous results if all the $t$'s are taken equal.\cite
{MullJeon},\cite{LL} We find 
\begin{eqnarray}
(\sigma |\hat{I}_{p}|\sigma ^{\prime })&=&\frac{\pi}{\hbar }\int
d\mathbf{p%
}_{1}d\mathbf{p}_{2}d\mathbf{p}_{3}\delta (\mathbf{p}_{1}+\mathbf{p}_{2}-%
\mathbf{p}_{3}-\mathbf{p}_{4})\delta (\epsilon_{\mathbf{p}_{1}}+\epsilon _{%
\mathbf{p}_{21}}-\epsilon _{\mathbf{p}_{31}}-\epsilon _{\mathbf{p}_{4}})
\nonumber \\&&%
 \sum_{\sigma _{2}}\left\{ -t_{\sigma \sigma _{2}}^{2}\left[ (n_{\mathbf{p}%
_{1}})_{\sigma \sigma ^{\prime }}(n_{\mathbf{p}_{2}})_{\sigma
_{2}\sigma _{2}}+\eta (n_{\mathbf{p}_{2}})_{\sigma \sigma
_{2}}(n_{\mathbf{p}%
_{1}})_{\sigma _{2}\sigma ^{\prime }}\right] \right.\nonumber \\&&
-t_{\sigma ^{\prime }\sigma _{2}}^{2}\left[
(n_{\mathbf{p}_{1}})_{\sigma
\sigma ^{\prime }}(n_{\mathbf{p}_{2}})_{\sigma _{2}\sigma _{2}}+\eta (n_{%
\mathbf{p}_{1}})_{\sigma \sigma _{2}}(n_{\mathbf{p}_{2}})_{\sigma _{2}\sigma
^{\prime }}\right] \nonumber \\&&
+\Bigl. 2t_{\sigma \sigma _{2}}t_{\sigma ^{\prime }\sigma _{2}}\left[ (n_{%
\mathbf{p}_{3}})_{\sigma \sigma ^{\prime }}(n_{\mathbf{p}_{4}})_{\sigma
_{2}\sigma _{2}}+\eta (n_{\mathbf{p}_{3}})_{\sigma \sigma _{2}}(n_{\mathbf{p}%
_{4}})_{\sigma _{2}\sigma ^{\prime }}\right] \Bigr\}  \label{collision}
\end{eqnarray}

We will linearize the kinetic equation for $\mathbf{m}_{p}$ around the
global equilibrium value $m_{p}^{(0)}\mathbf{\hat{z}}$ and use a
moment approach to compute the spin wave and diffusive damping just as
done previously.\cite{WillNik},%
\cite{Nikuni2} As in Ref.~\onlinecite{WillNik} we assume that the effective
longitudinal field $\Omega _{0}$ can be adjusted experimentally to zero. The
linearized longitudinal and transverse equations are 
\begin{equation}
\frac{\partial \delta m_{pe}}{\partial t}+\sum_{i}\left[ \frac{p_{i}}{m}%
\frac{\partial \delta m_{pe}}{\partial r_{i}}-\frac{\partial U}{\partial %
r_{i}}\frac{\partial \delta m_{pe}}{\partial p_{i}}\right] =\sum_{\sigma
}\sigma (\sigma |\hat{L}_{p}|\sigma )  \label{mlong}
\end{equation}
and 
\begin{eqnarray}
&&\frac{\partial \delta m_{p+}}{\partial t}+i\eta t_{12}\left(
m_{p}^{(0)}\delta M_{+}-M_{0}\delta m_{p+}\right)   \nonumber \\
&&+\sum_{i}\left[ \frac{p_{i}}{m}\frac{\partial \delta m_{p+}}{\partial r_{i}%
}-\frac{\partial U}{\partial r_{i}}\frac{\partial \delta m_{p+}}{\partial %
p_{i}}\right] =2(2|\hat{L}_{p}|1).  \label{mtrans}
\end{eqnarray} where $\hat{L}_{p}$ is the linearized form of 
$\hat{I}_{p}$

In the following, for brevity, we compute only results for the monopole and dipole
modes although experiments have detected the quadrupole modes. Similar 
arguments hold for the quadrupole case, which we will present in a 
longer publication.

\begin{figure}[h]
\centering
\includegraphics[width=5in]{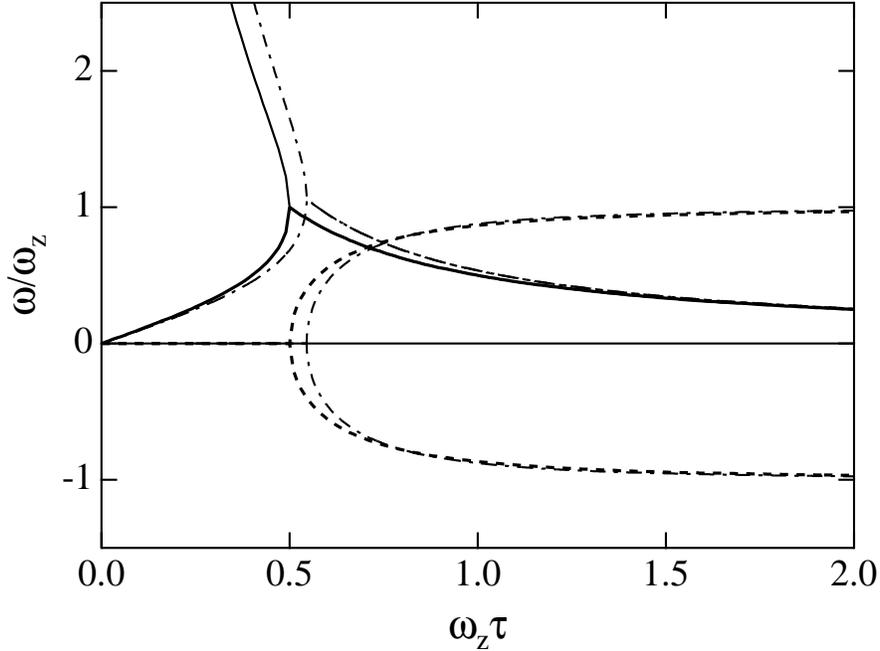} 
\caption{Real (dashed) and imaginary (solid) components of longitudinal
dipole spin wave modes versus average relaxation time $\tau_{\Vert}$. Note
the linear dependence of Im$(\omega)$ for small $\tau_{\Vert}$
characteristic of diffusive behavior. The dash-dotted lines represent the 
results of a numerical calculation to be discussed below.}
\label{fig:longdip}
\end{figure}

\emph{Longitudinal case}: We use a variational function of the form 
\begin{equation}
\delta m_{pz}=(a_{0}+a_{1}z\mathbf{+}a_{2}p_{z}\mathbf{)}m_{p}^{(0)}
\label{delmform}
\end{equation}
and take the $1,$ $z,$ and $p_{z}$ moments of the kinetic equation in both
the longitudinal and transverse cases. The results for the longitudinal
case, if we assume a time dependence of $\exp (i\omega t)$ for $a_{1}$ and $%
a_{2},$ are 
\begin{eqnarray}
da_{0}/dt &=&0  \label{a0eq} \\
i\omega a_{1}-\omega _{z}a_{2} &=&0 \\
i\omega a_{2}+\omega _{z}a_{1} &=&-\gamma_{\Vert}a_{2}  \label{LongaEqs}
\end{eqnarray}
with $\gamma _{\Vert}=4\gamma _{0}/3$ where $\gamma _{0}=\pi \beta m^{3}\bar{%
\omega}^{3}t_{12}^{2}N/h^{4}$ comes from integrating the collision
integral. Eq.~(\ref{a0eq}) indicates that the monopole mode does not decay in the longitudinal case, which is consistent with the conservation of
magnetization. 
The second line is the magnetization equation of continuity. The relaxation
rate $\gamma _{\Vert}$ agrees with that derived in Ref.~\onlinecite{WillNik}.
The dipole spectrum is plotted in Fig.~\ref{fig:longdip} as a function of $\tau
_{_{\Vert }}\equiv 1/\gamma _{\Vert}$, the \emph{spatially averaged} collision
time. In the small $\tau _{_{\Vert }}$ limit, one finds 
\begin{equation}
\omega _{\parallel }=i\omega _{i}^{2}\tau _{_{\Vert }},
\end{equation}
which has the form of the lowest-order solution of a diffusion equation in a
harmonic potential. 

\emph{Transverse case}: We again use the form of Eq.~(\ref{delmform}).
Taking $1,$ $z,$ and $p_{z}$ moments of Eq.~(\ref{mtrans}) yields the results 
\begin{eqnarray}
da_{0}/dt &=&-\gamma _{T}a_{0}  \label{DipTrans1} \\
i\omega a_{1}-\omega _{z}a_{2} &=&-\frac{1}{2}\gamma _{T}a_{1}  \label{cont}
\\
i(\omega -\omega _{M})a_{2}+\omega _{z}a_{1} &=&-\gamma _{\perp}a_{2}
\label{DipTrans}
\end{eqnarray}
where 
$\gamma _{T}=\gamma _{0}(1+\eta )\sum_{\sigma }\left( \frac{t_{\sigma \sigma
}-t_{12}}{t_{12}}\right) ^{2}f_{\sigma }$
with $f_{\sigma}=N_{\sigma}/N,$ and 
\begin{equation}
\gamma _{\perp}=\gamma _{\Vert}\left[ \frac{7R-3S}{8t_{12}^{2}}\right]
\label{gamma2}
\end{equation}
with $R=(1+\eta )\sum_{\sigma }t_{\sigma \sigma }^{2}f_{\sigma }+(1-\eta
)t_{12}^{2}$ and 
$S=2t_{12}[(1+\eta )\sum_{\sigma }t_{\sigma \sigma }f_{\sigma }-\eta 
t_{12}],$
and the mean-field frequency is 
\begin{equation}
\omega _{M}=\eta \frac{t_{12}\mathcal{M}}{\hbar }\left( \frac{\beta \hbar 
\bar{\omega}}{\sqrt{2}\lambda }\right) ^{3}
\end{equation}
where $\lambda$ is the thermal wavelength.

\noindent Comments:

1) If the interactions parameters $t_{ij}$ are all equal, we have
$\gamma _{T}=0,$ $R=S=2t^{2}$ so that $\gamma _{\perp}=\gamma _{\Vert}.$
Eqs.~(\ref {DipTrans1})-(\ref{DipTrans}) then reduce to those of
Ref.~\onlinecite{WillNik} and the longitudinal and transverse
relaxation rates are the same, which agrees with the standard result
for a homogeneous real spin system in the Boltzmann limit.

2) For fermions, we have $\eta =-1$, so that, even if the $t$'s are \emph{not%
} equal, $\gamma _{T}=0$ and $\gamma _{\perp}=\gamma _{\Vert}.$

3) For bosons with unequal $t$'s, the spatial averaged transverse
relaxation rate is not generally the same as the longitudinal. 
Moreover, we have a $T_{2}$-type relaxation rate for $a_{0}$ and in
the equation of continuity (\ref{cont}).  The interaction anisotropy
behaves something like a dipole-dipole interaction allowing relaxation
of the transverse spin, an effect noted previously in
Ref.~\onlinecite{Nikuni2}.

If, for now, we take $\gamma _{T}=0,$ then the lowest mode in the
hydrodynamic limit takes the form 
\begin{equation}
\omega _{\bot }=\frac{\omega _{z}^{2}(i-\mu \mathcal{M)}\tau _{\bot }}{%
\left[ 1+(\mu \mathcal{M)}^{2}\right] }.  \label{LRfreq}
\end{equation}
where $\tau _{\bot }\equiv 1/\gamma _{\perp}$ and the so-called ``spin-rotation
parameter'' $\mu=\omega_{M}\tau_{\perp}.$ 
The form of Eq.~(\ref{LRfreq}) is the hydrodynamic frequency as modified by
spin rotation. \cite{cand},\cite{MullJeon},\cite{LL}. The first term is the
effective diffusion frequency while the second is the dipole-mode
pseudo-spin-wave frequency.

\begin{figure}[h]
\centering
\includegraphics[width=5in]{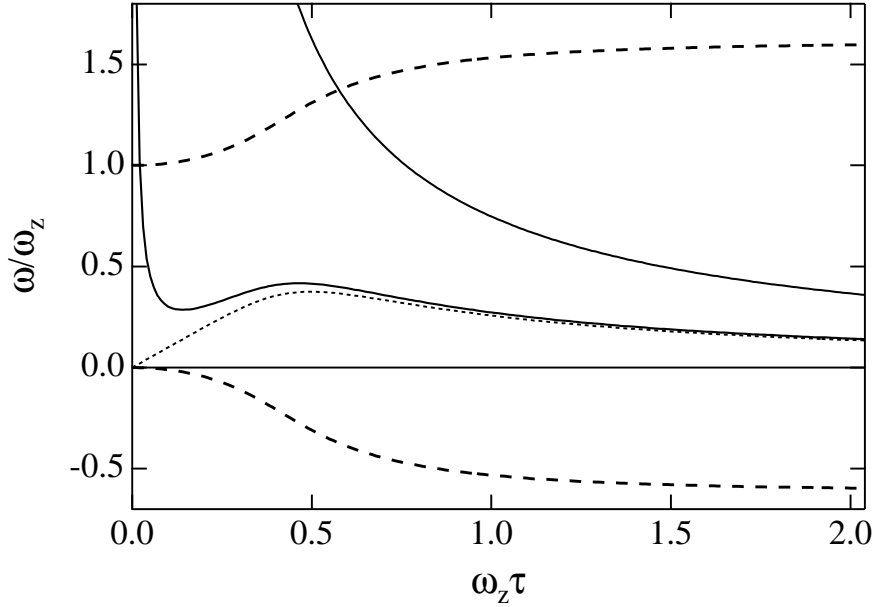} 
\caption{Real (dashed) and imaginary (solid) components of transverse dipole
spin wave modes versus average relaxation time $\tau_{\perp}$ when the
transverse spin is not conserved. The dotted line shows the lower imaginary
mode when the transverse decay rate $\gamma_{T}=0$. The mean-field frequency 
$\omega_{M}$ is taken as $\omega_{z}$. The linear behavior of Im$(\omega)$
for small $\tau_{\perp}$ characteristic of diffusive behavior is destroyed
and replaced by a divergence within the moments method used here.}
\label{fig:transdip}
\end{figure}

The effect of non-zero $\gamma _{T}$ is to allow a $T_{2}$ relaxation
of the transverse spins.  The results are shown in Fig.~2, where we
have set $1/\tau _{\perp }=\gamma _{\Vert}$, $\gamma _{T}=0.02/\tau
_{\perp }, $ and $\omega _{M}=\omega _{z}.$ In the small $\tau _{\perp
}$ limit, one no longer has the hydrodynamic decay rate approaching
zero, but instead it diverges at the origin because Im$(\omega
)\approx \left( \omega _{z}^{2}\tau _{\perp }+\gamma _{T}\right)$, and
$\gamma _{T}\sim 1/\tau _{\perp }$.  However, although suitable for
finite $\omega \tau$, the moments method is inadequate in the
hydrodynamic limit.  It fails because the simple forms
assumed for spatial dependence cannot adjust to the spatially
dependent relaxation rates.  One must solve local equations
numerically for the spatial behavior.

To obtain the hydrodynamic equations we expand the momentum
distribution in terms of Hermite polynomials
\begin{equation}
\delta m_{pz}= e^{-\beta p^2/2m}\sum_{k=0} c_k(z,t)H_k(p)
\label{R1}
\end{equation}
Substituting this into the kinetic equations, integrating
over the momentum, and keeping terms lowest order in $\tau_{\perp}$ gives, in the transverse case,

\begin{eqnarray}
\partial_t \delta M_++\partial_z  J_+ =-\gamma_T(z) \delta M_+\\
\partial_t J_++\frac{kT}{m}\partial_z \delta M_++ \omega_z^2z \delta
 M_+&+&i\omega_M(z)J_+ 
 \approx-\gamma_{\perp}(z)J_+
\label{R4b}
\end{eqnarray}
where $\delta M_+(z,t)=c_0(z,t)$ is the nonequilibrium magnetization density,
$J(z,t)=\int d\mathbf{p}/h^3 (p/m)\delta m_{pz}= c_1(z,t)$ is the spin
current, and $\gamma_{\perp}(z)=\gamma_{\perp}(0)\exp(-\beta
m\omega_z^2 z^2/2)$. Analogous equations hold in the longitudinal 
case. On the RHS of Eq.~(\ref{R4b}) the $k=1$ momentum
distribution has been treated as an eigenfunction of the linearized
collision integral.  This is justified by a numerical calculation of
the matrix elements of the collision integral, which gives
\begin{eqnarray}
&&L_{\perp}[H_1(p)] = - \gamma_{\perp}(0)(1.000H_1(p) +
0.123H_3(p)\nonumber\\
&&-0.00094H_5(p)+ ...) \approx -\gamma_{\perp}(0)H_1(p)
\label{R3}
\end{eqnarray}

The eigenvalues of the hydrodynamic equations have been calculated
numerically for the dipole mode with boundary conditions $\delta
M(0)=0$, $J(0)=1$, and $J(\infty)=0$, and the monopole mode with
boundary conditions $\delta M(0)=1$, $J(0)=0$, and $J(\infty)=0$.  For
the longitudinal and isotropic transverse cases this leads to only
small corrections to the $\tau \rightarrow 0$ part of the spectra
obtained by the moments method as shown in Fig.~\ref{fig:longdip}.

 \begin{figure}[h] 
 \centerline{\includegraphics[width=5in]{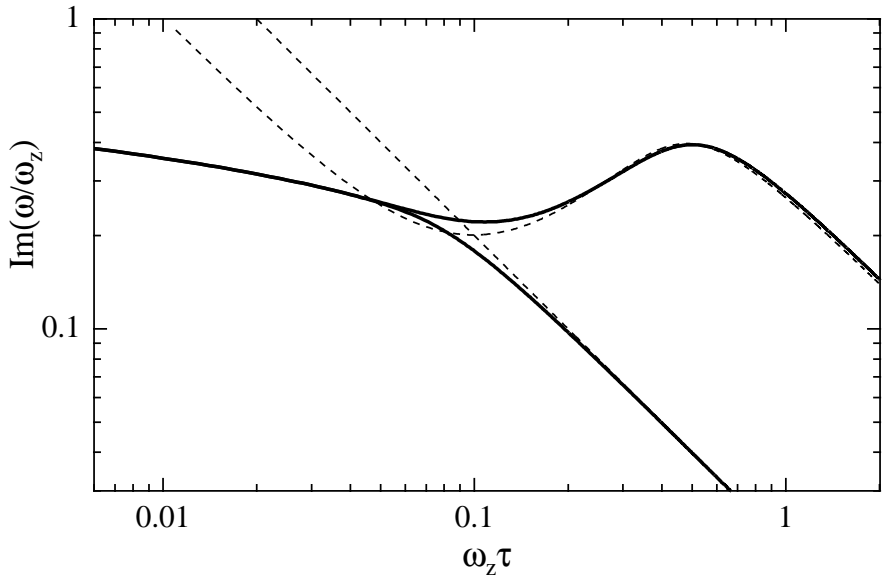}} 
 \caption{Imaginary part of the spin-wave spectrum vs.
 $\omega_z\tau_{\perp}$ for the monopole and dipole modes with
 $\omega_{M}= \omega_{z} $, $\gamma_T=0.02\gamma_{\perp}$, and
 $\gamma_{\perp} \approx \gamma_{\parallel}$ for both the moments
 method (thick) and hydrodynamic calculations (dashed).   }
 \label{fig3} 
 \end{figure}

 \begin{figure}[h] 
 \centerline{\includegraphics[width=5in]{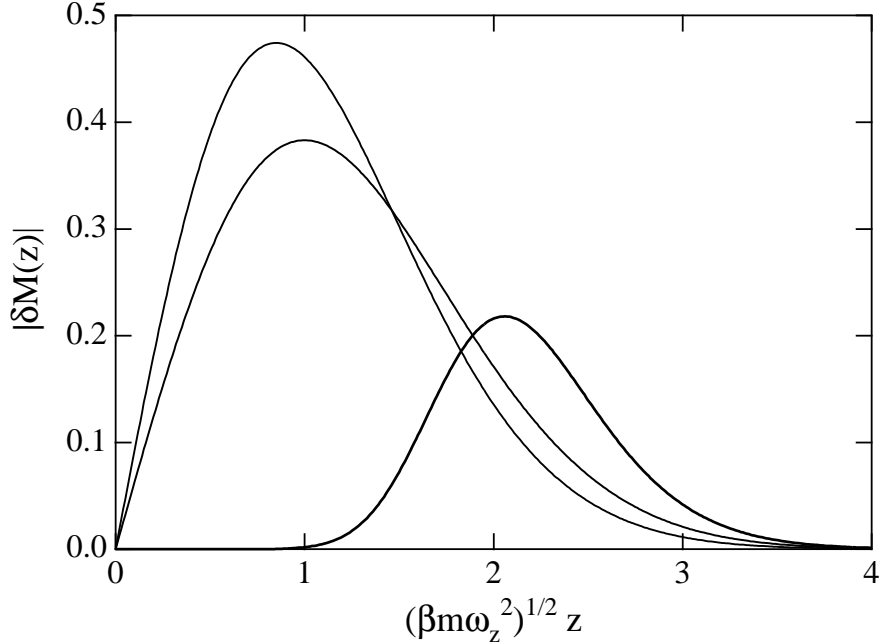}} 
 \caption{Profiles $|\delta M(z)|$ of the the dipole modes
 vs.  $\sqrt{\beta m\omega_z^2}z$, normalized with a Hermite weighting function.  Tallest peak is the hydrodynamic
 mode for $\gamma_T=0$ and $\tau_{\perp}=0.01$.  Middle peak is the
 moments method ansatz.  Smallest peak is the hydrodynamic mode for
 $\gamma_T=0.02\gamma_{\perp}$ and $\omega_z\tau = 0.01$.  }
 \label{fig4} 
 \end{figure}

However, for $\gamma_T >0$ the hydrodynamic spectrum differs
qualitatively from that of the moments method calculation.
As $\tau_{\perp}\rightarrow 0$ the hydrodynamic dipole and monopole modes
do not decay at a rate $\sim 1/\tau_{\perp}$, but instead decay at a
slower rate $\sim \sqrt{\log{(1/\omega_z\tau_{\perp})}}$ (See Fig.~\ref{fig3}.)
In fact, at small enough $\omega_z\tau_{\perp}$ the $T_2$ decay of the
magnetization at the center of the trap causes the monopole and dipole
modes to coalesce into spin-waves localized on the lower density
regions on the left and right sides of the trap. (See
Fig.~\ref{fig4}.) 

In experiments on Rb, the interaction anisotropy is very small.  To
test the novel effects predicted here one might use Na,\cite{sodium}
which has a difference in interaction paraments; Numerically we
estimate that for $^{23}$Na $\gamma_{\perp}$ can differ from
$\gamma_{\perp}$ by as much as 14\% with
$\gamma_{T}/\gamma_{\perp}\approx 0.04$. Interaction differences might
also be induced by using Feshbach resonance methods.

We thank Dr.\ Jean-No\"{e}l Fuchs and Prof. David Hall for useful
discussions.


\begin{thebibliography}{99}
\bibitem{footnote} We consider temperatures such that the distribution 
function is Boltzmann, but the pair interactions are quantum mechanical 
with s-wave scattering predominating. This can occur if the 
interparticle spacing is much greater than the de Broglie wavelength 
which, in turn, is much greater than the scattering length.
\end{thebibliography}
\end{document}